%% file: paper.tex
\documentclass{llncs}

\usepackage{cite}
\usepackage{amsmath,amssymb,amsfonts}
\usepackage{graphicx}
\usepackage{subfigure}
\usepackage{textcomp}
\usepackage{xcolor}
\usepackage{enumitem}
\usepackage{siunitx}
\usepackage{algorithm}
\usepackage{algpseudocode}

\usepackage{listings}
\usepackage{color}
\definecolor{gray}{rgb}{0.4,0.4,0.4}
\definecolor{darkblue}{rgb}{0.0,0.0,0.6}
\definecolor{cyan}{rgb}{0.0,0.6,0.6}
\lstdefinelanguage{XML}
{
  morestring=[b]",
  morestring=[s]{>}{<},
  morecomment=[s]{<?}{?>},
  stringstyle=\color{black},
  identifierstyle=\color{darkblue},
  keywordstyle=\color{cyan},
  morekeywords={xmlns,version,type}
}

\newcommand{\subscript}[2]{$#1 _ #2$}
\newcommand{\paragraphb}[1]{\vspace{0.03in}\noindent{\bf #1} }

\makeatletter
\newcommand{\algmargin}{\the\ALG@thistlm}
\makeatother
\newlength{\whilewidth}
\algdef{SE}[parWHILE]{parWhile}{EndparWhile}[1]
  {\parbox[t]{\dimexpr\linewidth-\algmargin}{%
     \hangindent\whilewidth\strut\algorithmicwhile\ #1\ \algorithmicdo\strut}}{\algorithmicend\ \algorithmicwhile}%
\algnewcommand{\parState}[1]{\State%
  \parbox[t]{\dimexpr\columnwidth-\algmargin}{\strut #1\strut}}

\algnewcommand{\LineComment}[1]{\State \(\triangleright\) #1}

\begin{document}

\title{
	A Unified Access Control Model for Calibration Traceability in Safety-Critical IoT
	\thanks{
		The authors would like to acknowledge the support of the National Measurement System of the UK Department of Business, Energy \& Industrial Strategy, which funded this work as part of NPL's Data Science program.
	}
}

\author{
	Ryan Shah, Shishir Nagaraja
}
\institute{University of Strathclyde, Glasgow}

\maketitle

\thispagestyle{plain}
\pagestyle{plain}

\begin{abstract}

Accuracy is a key requirement of safety-critical IoT (SC-IoT) systems.
Calibration plays an important role in ensuring device accuracy within an
IoT deployment. The process of calibration involves a number of parties
such as device users, manufacturers, calibration
facilities and NMIs. These parties must collaborate to support calibration. Calibration checks often precede safety-critical operations such as preparing a robot for surgery, requiring inter-party interaction to complete checks. At the same time, the parties involved in a calibration ecosystem may share an adversarial relationship with a subset of other parties. For instance, a surgical robot manufacturer may wish to hide the identities of third-parties from the operator (hospital), in order to maintain confidentiality of business relationships around its robot products. Thus, information flows that reveal \textit{who-calibrates-for-whom} need to be managed to ensure confidentiality. Similarly, meta-information about \textit{what-is-being-calibrated} and \textit{how-often-it-is-calibrated} may compromise operational confidentiality of a deployment. For example, calibration-verification of connected medical devices may reveal the timing of surgical procedures and potentially compromise PII when combined with other meta information such as patient admission and exit times.
We show that the challenge of managing information flows
between the parties involved in calibration cannot be met by any of the
classical access control models, as any one of them or a simple
conjunction of a subset such as the lattice model fails to meet the
desired access control requirements. We demonstrate that a new unified
access control model that combines BIBA, BLP, and Chinese Walls holds rich
promise. We study the case for unification, system properties, and develop an
XACML-based authorisation framework which enforces the unified model.
We show that upon evaluation against a baseline simple-conjunction of the
three models individually, our unified model outperforms nearly two-fold
and demonstrates it is capable of solving the novel access control
challenges thrown up by digital-calibration supply chains.


\end{abstract}


\input{sections/introduction.tex}
\input{sections/background.tex}
\input{sections/model.tex}
\input{sections/evaluation.tex}
\input{sections/discussion.tex}
\input{sections/related_work.tex}
\input{sections/conclusion.tex}

\bibliographystyle{splncs04}
\bibliography{references}

\end{document}

%% file: sections/introduction.tex
\section{Introduction}
\label{sec:intro}

Many safety-critical IoT (SC-IoT) systems, such as surgical robots, are employed to address the need for higher levels of accuracy and precision. For example, the Robodoc surgical robot has shown a decrease in the number of complications in hip surgeries by providing higher accuracy for implant sizing and positioning within the bone compared to traditional surgery~\cite{bargar1998primary}; and in autonomous vehicles, a variety of sensor devices are heavily relied on to provide assistance for real-time decision making~\cite{yaugdereli2015study}. This has led to an increased desire in the adoption
of such systems in high-assurance environments, where the system can outperform
relative traditional methods with lowered risk of complications.

\begin{figure}
  \centering
  \includegraphics[width=0.75\linewidth]{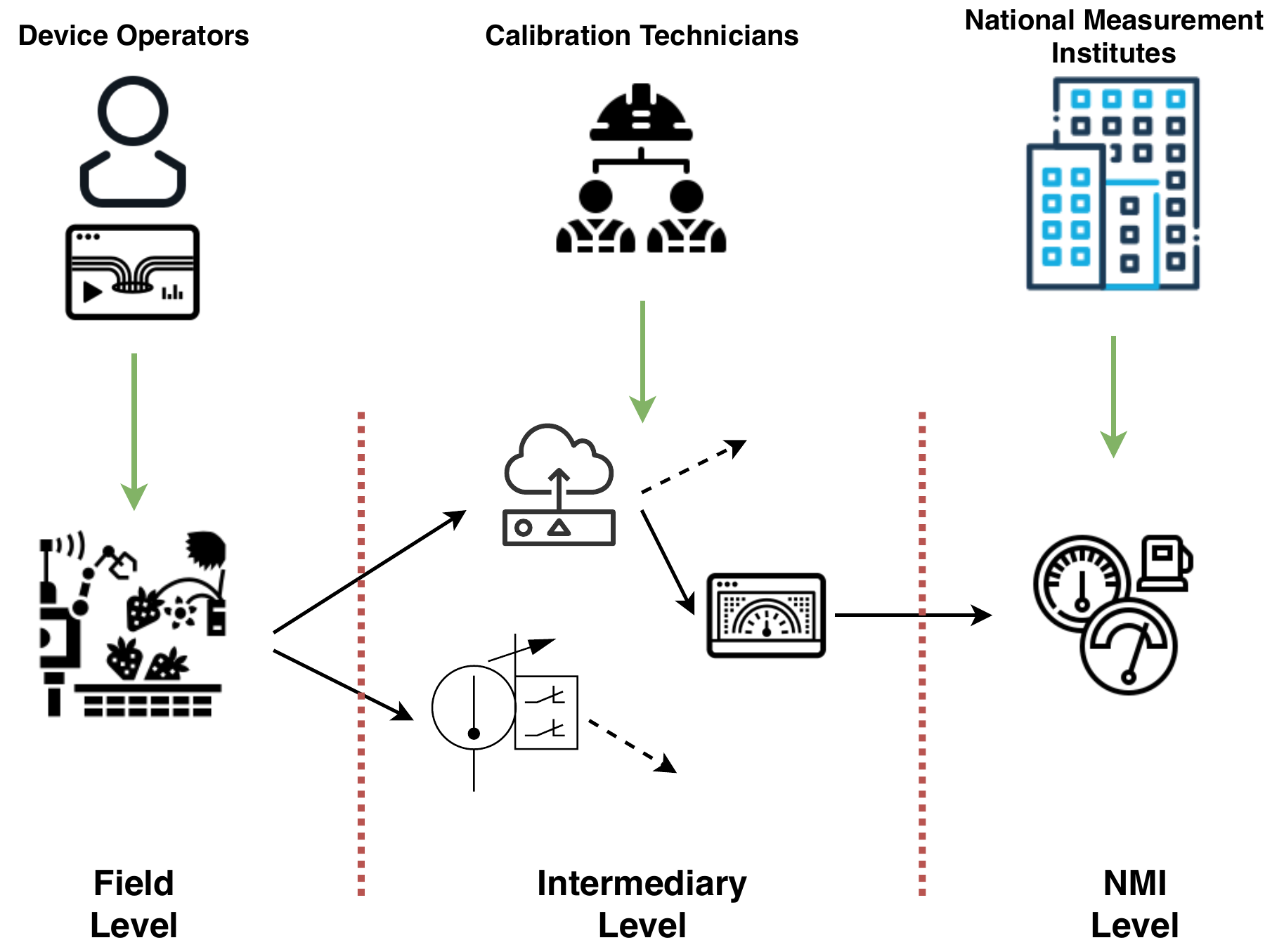}
  \caption{Calibration Hierarchy for an SC-IoT System}
  \label{fig:hierarchy}
\end{figure}

While IoT device security and privacy has received a lot of attention~\cite{hwang2015iot,abomhara2014security,dabbagh2019internet}, the security of calibration
has received little attention. Calibration processes involve many parties. These parties form a hierarchy of other calibrated devices within which they
interact (Figure~\ref{fig:hierarchy}). First, at the field level we have device operators who
control, interact with and maintain the system and its components. Second, we have
one or more intermediary calibration facilities who employ technicians
which interact with other calibrated devices within the hierarchy. Finally, we
have National Measurement Institutes (NMIs) at the highest level, such as NIST
(USA) and NPL (UK), who maintain homeostatic control within the hierarchy.


While many of these parties do cooperate in harmony, for example NMIs who
manage publicly available calibration information and act as the root of trust,
a subset of them may share an adversarial relationship (in direct competition
with one another). For example, Uber holds a fleet of autonomous cars which
employ a number of sensors (i.e. LIDAR, infrared (IR), RADAR, etc.) that
require calibration to make sure that all sensed data is accurate and reliable.
Some of these devices may be calibrated by a third-party calibration service
provider (i.e. an intermediary facility) who they may wish to remain
confidential to show other parties (i.e. business partners) that they are
\textit{responsible} for their own calibration, helping to maintain the business
relationships surrounding its autonomous fleet. By revealing this information to
parties in competition, the confidentiality surrounding among the participants
(\textit{who-calibrates-for-whom}) would be compromised. Furthermore, although
protecting the confidentiality of business relationships surrounding parties
with conflicts-of-interest, the information regarding what devices are
calibrated and how frequent these calibration processes are carried out can
also compromise the operational confidentiality of SC-IoT deployments. For
instance, if we consider a technician from a third-party calibration facility
who calibrates for Uber, to calibrate devices for autonomous vehicles at
another organisation (i.e. Lyft), information about what devices are being
calibrated at Uber and internal calibration processes could be leaked to them.
Similarly, by monitoring the calibration processes for sensors in autonomous
vehicles, it is possible to map the operational usage of these components. For
instance, by monitoring calibration-verification of LIDAR sensors, collecting
the calibration offsets and other meta-data, it is possible to reveal how these
sensors are used.

Ultimately, we observe that the potential compromise of calibration integrity
and confidentiality, and the adversarial relationships (conflict) between
interacting parties in the calibration ecosystem, present
us with a unique set of information flows where meta-information such as what
is calibrated, how often it is calibrated and who calibrates for who, need
to be managed. Importantly, a solution that can manage these information
flows requires all stakeholders to work together, presenting us with an
interesting access control challenge. While existing access control models
such as Chinese Walls may provide some benefit, we note that individual models
or a combination of such would fail to control the information flows. Therefore,
an effective solution to address this should satisfy the following requirements:

\begin{enumerate}[label=(\subscript{R}{{\arabic*}})]
	\item How can we manage the adversarial relationships between a subset of
		  interacting parties in the calibration ecosystem, to avoid unintended
		  disclosure of information?
	\item How can we protect the confidentiality of business relationships
		  (\textit{who-calibrates-for-whom}) whilst providing transparency
		  for calibration processes?
	\item How can we ensure operational confidentiality of SC-IoT deployments
		  by protecting \textit{what-is-being-calibrated} and
		  \textit{how-often-it-is-calibrated}?
	\item As many SC-IoT systems are time-critical, how can we support
		  calibration processes to be carried out efficiently (\textit{on-the-fly}) whilst ensuring \subscript{R}{1}--\subscript{R}{3}?
\end{enumerate}

In this paper, we propose an access control model that unifies the
BIBA, BLP and Chinese Walls models which satisfies the above requirements.
The rest of the paper is organised as follows. In Section~\ref{sec:background}, we further detail the inadequacies of
the current state-of-the-art in calibration and access control, and discuss
the observed information flow constraints which demands a novel
unification of BIBA, BLP and Chinese Walls. Following this, we detail
our unified access control model in Section~\ref{sec:model} and practically
evaluate its performance using an attribute-based authorisation mechanism
in Section~\ref{sec:evaluation}. In Section~\ref{sec:discussion}, we
provide a discussion of the presented work, provide
related work in Section~\ref{sec:related} and conclude in Section~\ref{sec:conclusion}.

%% file: sections/background.tex
\section{Calibration Traceability and Access Control: A Case for Unification}
\label{sec:background}

A key requirement for SC-IoT systems is to ensure that all components that make
up such a system are highly accurate, with very low margin for error. While all
aspects of security regarding SC-IoT is important, such as communication and
control plane security~\cite{wang2010security,hasan2019protecting}, the
importance of calibration has been misunderstood.

\begin{figure}
  \centering
  \includegraphics[width=0.75\linewidth]{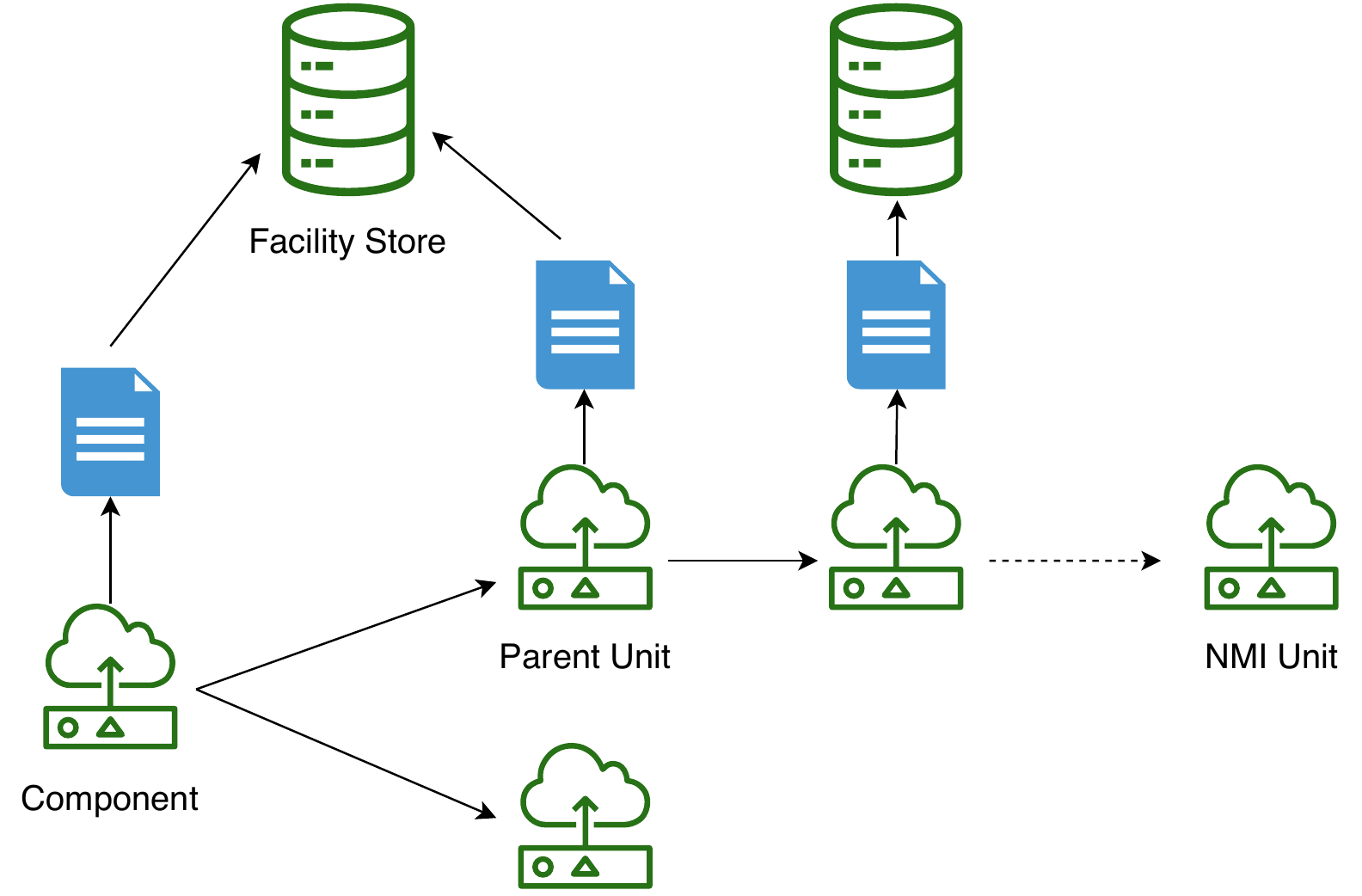}
  \caption{Traceability Chain}
  \label{fig:traceability}
\end{figure}

Calibration is a key factor which contributes to the accuracy and reliability of
our devices, such as the readings produced from sensing equipment. Simply put,
the process of calibration is a comparison of a given device, such as a sensor
in an SC-IoT system, with a more accurate (parent) device. Specifically, the
accuracy and reliability of our device's output is derived from its parent.
However, while having our devices being calibrated is important, the validity
of its calibration is key. An important requirement for valid calibration is
having an unbroken chain of traceable calibration to national
standards~\cite{vim2004international,de2000calibration} (Figure~\ref{fig:traceability}).
This means that valid calibration requires a complete trace of calibration, starting
with a comparison between our device and a more accurate parent device,
who will then be compared against an even more accurate device, and
so forth until we compare against a device that is as accurate as possible
(where the accuracy and uncertainty of output is level with national standards).

With the considerations of SC-IoT systems, we observe several inadequacies with
the current state-of-the-art in calibration traceability.

First, current calibration traceability processes are inefficient. As an output
of calibration, a report (hereafter refered to as a calibration report) is
produced, which details the information used in traceability verification
(such as calibration offsets, uncertainty measurements, operating ranges, etc.).
Furthermore, these are stored centrally within the organisation who carried out the calibration of
the associated device. For example, a device calibrated by a technician
at some intermediary calibration facility will typically have its report stored
locally. To carry out verification checks (Figure~\ref{fig:traceability}),
a subject (device owner, technician performing calibration, etc.)
must first request the device's report from the holding organisation.
Once access is granted, the parent units of the device can be identified
from the report and the reports for these devices must also be accessed,
and so forth to national standards. While the process is relatively trivial to
conduct, many SC-IoT systems are comprised of large numbers of devices, each of
them having their own individual traceability chains. With this in mind, a
verification process which ensures there is a complete chain of calibration for
a single device could take a few hours but for an entire SC-IoT system,
verification time would increase significantly.

Second, given that calibration reports are stored in different organisations
for a single device's traceability chain, each organisation could have their
own security requirements to grant access to reading a report. For example, test
and measurement devices used to calibrate military equipment could have their
reports classed as confidential, or perhaps the organisation who owns the
equipment wishes to remain secret to outsiders). Therefore, by verifying
calibration along each step in a device's chain could leak confidential information
about how the device is used.

Third, while nonsensical, it is entirely possible that a technician at the field
level in the chain could fabricate reports for (potentially non-existent)
devices further up the chain, such as for intermediary calibration facilities
or National Measurement Institutes (NMIs). As the calibration status
of downstream equipment is dependent on upper levels, the integrity of
calibration traceability and ultimately the device itself could be
compromised. Furthermore, it is important to allow subjects conducting
traceability verification checks at lower levels (i.e. the field level)
to access reports at each upper level for completeness, while disallowing
those operating at the field level to write reports for upper levels,
and vice-versa.

Finally, the potential for conflicts of interest between competing calibration
facilities arise. For example, a calibration technician who calibrates devices
for one organisation, should not be allowed to calibrate devices for another,
to avoid leaking potentially sensitive information about the device and the
organisation's internal calibration processes.

Upon observation of the problem space surrounding the current
state-of-the-art, and the considerations surrounding SC-IoT, we can see
that the key failures are provoked by varying access and security requirements
for verifying calibration traceability within a multi-level hierarchy.
Therefore, it is important to devise an appropriate solution that is not
only highly efficient, but also robust to adversaries who target the
confidentiality and integrity of such systems from the calibration
attack vector, and the competitive nature of actors within the calibration
ecosystem.

\subsection{Information Flow Constraints}

\begin{figure}
  \centering
  \includegraphics[width=0.6\linewidth]{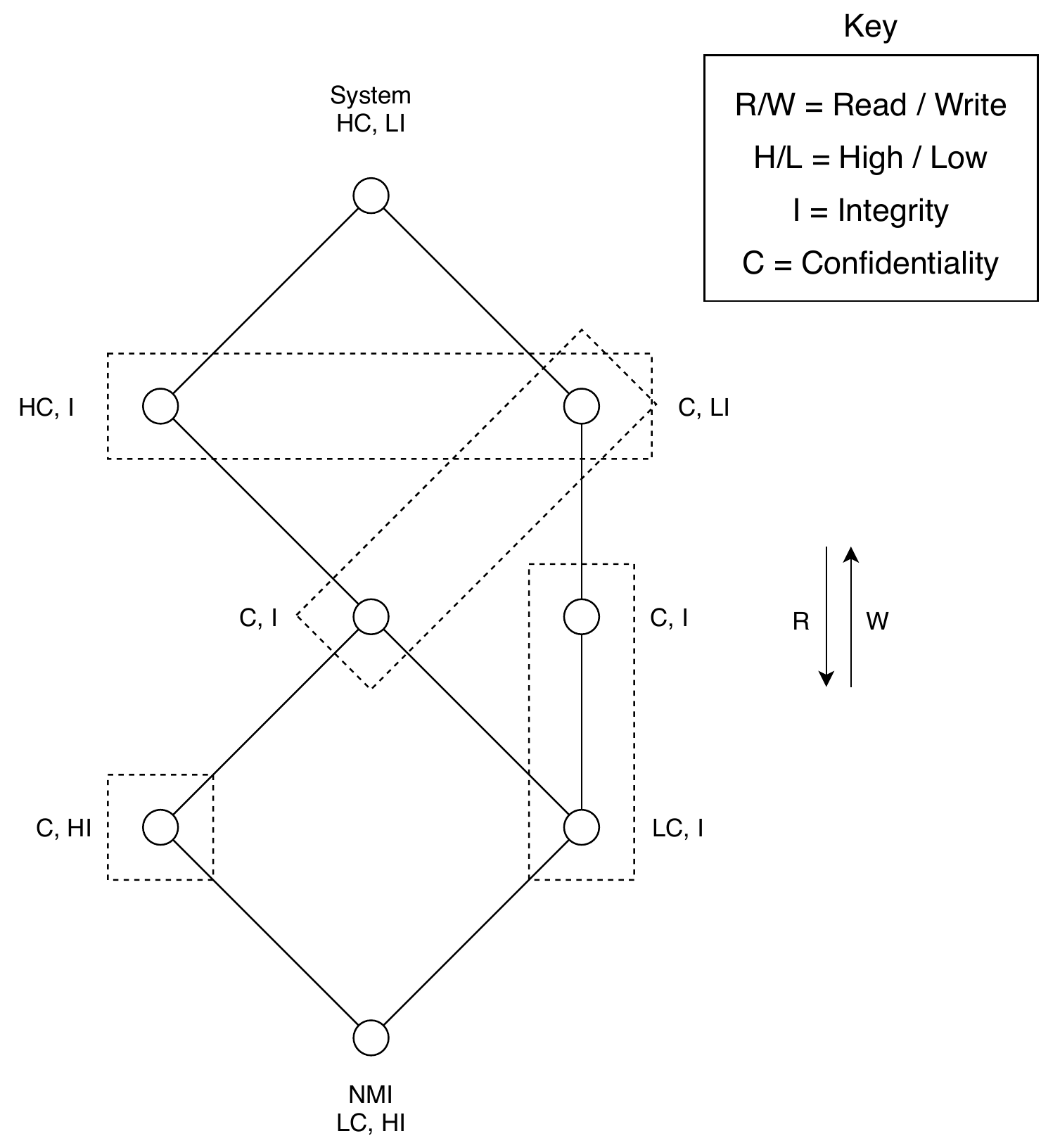}
  \caption{Information Flow Model}
  \label{fig:informationflowmodel}
\end{figure}

An important requirement for designing an effective solution that meets
our requirements is to define the information flow constraints which should be
enforced. If we design the solution well, coinciding with the multi-level hierarchy
that is natural to traceability chains, we can ultimately reduce the
number of critical components by an order of magnitude. For subsequent
discussion, we refer to the information flow model depicted in
Figure~\ref{fig:informationflowmodel}.

\subsubsection{Multi-Level Integrity}

Our first requirement is to maintain the integrity of calibration. By
maintaining the integrity of calibration reports at the root
level, we can reduce the damage inflicted to intermediate levels between
the root level and the components which make up a safety-critical system.
As a reminder, if compromise occurs at the root level in a traceability
chain, then the validity of all subsequent levels towards the field level
is put into question. By maintaining the integrity of calibration,
we can limit any damage to its immediate locality, as opposed to
inflicting widespread damage.

Overall, we observe that with respect to traceability, information
flows from a high integrity source (the root calibration units at NMIs)
to a low integrity destination (the components at the field level).

\subsubsection{Multi-Level Confidentiality}

The second of our requirements is to maintain confidentiality among
participants whilst enabling transparency in traceability chains.
Interestingly, we observed that it is relatively trivial to map
components in traceability chains to real actors. For example, a
component's calibration report could reveal the facility at which
calibration was carried out, as well as information that denotes
a facility's internal calibration processes. With consideration to NMIs,
the information is globally available, so in this case the mapping of
equipment here does not succumb to confidentiality concerns. However,
in the case of intermediate levels, who do not wish to reveal information
to other levels above or below them, the possibility of such a mapping
becomes of real concern.

Furthermore, the timing of traceability verification checks could reveal
information about system operations. For example, in the context of
surgical robots, the timing of surgical procedures could be leaked as a
result of monitoring verification checks and when combined with
other sources of information such as patient admission and exit times,
a violation of patient confidentiality is possible. As well as this,
the verification process should not reveal information about the
deployment to intermediaries involved between the deployed system and
the public-facing NMI at the highest level. The reason for this is simple
- calibration traffic could leak information to parties further down the
chain. For example, a device manufacturer may employ the services of a
third-party calibration facility to calibrate the system's sensors, but
may not want to reveal who the third party is. Thus, the traceability
chain should be verifiable in a manner, such that information about the
setting where the system is deployed is not leaked to other facilities,
and likewise from these facilities to the NMIs.

Overall, we note that the field level where systems are deployed must
retain the highest level of confidentiality, whilst calibration
present at the public-facing root level have the lowest confidentiality
requirements.

\subsubsection{Conflicts of Interest}

Finally, the last requirement for an appropriate solution is to prevent
unintended disclosure of information, specifically relating to the nature
of conflicts between calibration providers and organisations that manage the
systems. For example, Uber may wish to hide the identity of the service provider that calibrates their LIDAR sensors for their driverless fleet. A
technician from this company should not be allowed to calibrate for other
providers of driverless vehicles, to avoid the leakage of sensitive business information to
competitors. As such, it is vital that we compartmentalise those in competition
into conflict of interest sets, such that this sensitive information cannot flow
between them.

%% file: sections/model.tex
\section{A Unified Access Control Model for Calibration Traceability}
\label{sec:model}

Upon review of the information flow constraints related to calibration
traceability, we note this presents us with a novel access control
challenge to which the unification of the BLP, BIBA and Chinese Wall
models fits well. However, the constraints assigned to information
flow defined by BIBA and BLP contradict each other. Fortunately, by
taking Sandhu's observation~\cite{sandhu1993lattice} that BLP and BIBA
are in fact the same model, we can see that our information flow
(Figure~\ref{fig:informationflowmodel}) is an instance of this observation and by reversing BIBA, we can relieve this contradiction.

In current calibration-related business practices, we identified that
these models are manually implemented. Within calibration facilities,
the BLP model is exercised on calibration reports. Specifically, some
calibration reports may have a degree of confidentiality, or the
technicians calibrating equipment may do so under a non-disclosure
agreement, and it is likely that internal systems and individuals
maintain these levels of confidentiality. As for the notions of
integrity handles by BIBA is inherent to the calibration ecosystem,
but is not rigorously enforced. Instead, it is implied that integrity
must be maintained at all levels, specifically at the root level,
but any problems are not uncovered until the annual cycle of traceability
verification, at which point the liabilities that arise due to invalid
calibration may not be settled as easily.

For subsequent discussion of the unification of the three models,
we define the terminology using notation based on the work of
Sandhu~\cite{sandhu1993lattice}.

\paragraphb{(Definition) Conflict of Interest Set} A Conflict of Interest (COI) set is defined as the set of subsets, where
each subset corresponds to the calibration providers in who have a direct
conflict of interest with each other. Following standard notation, we
denote the set of $n$ COI sets as $\{COI_1, COI_2, \ldots, COI_n\}$,
where each set $COI_i = \{1, 2, \ldots, m\}$ contains the set of $m$
technicians in conflict.

\paragraphb{(Definition) Set of Integrity Labels} The set of integrity labels is denoted as $\Omega = \{\omega_1, \omega_2, \ldots, \omega_q\}$,
where each label corresponds to a unique integrity level. Since we
reverse BIBA to match the information flow of BLP, each integrity level
also corresponds to a unique confidentiality level.

\paragraphb{(Definition) Security Label} A security label is defined as a set of two n-sized vectors $\{[i_1, i_2, \ldots, i_n], [p_1, p_2, \ldots, p_n]\}$, where $i_j \in \{COI_j \cup \perp \cup  T\}$, $p_j \in \Omega$ and $1 \leq j \leq n$.

\begin{itemize}
	\item Where $i_j = \perp$, the calibration traceability chain does not contain information from any provider in $COI_j$.
	\item Where $i_j = T$, the calibration traffic contains information from {\em at least} two facilities who are in a conflict of interest set $COI_j$.
	\item Where $i_j \in COI_j$, the calibration traffic contains information from the corresponding calibration facility in $COI_j$.
\end{itemize}

\paragraphb{(Definition) Dominance Relations} We define the (transitive) dominance relations between security labels as
follows, where the notation $l_j[i_k]$ denotes the $i^{th}_k$ element of
the label $l_j$. We say that a security label $l_1$ dominates a label
$l_2$, denoted by $l_1 \geq l_2$, where
\(l_1 \geq l_2 \iff \forall i_k,p_k = (1,2,\dots,n) [((l_1[i_k] = l_2[i_k]) \vee (l_2[i_k] = \perp) \vee (l_1[i_k] = T)) \wedge (l_1[p_k] \leq  l_2[p_k])]\).

\begin{itemize}
	\item A label $l_1$ dominates a label $l_2$, provided that $l_1$ and $l_2$ agree whenever $l_2$ is not public or in conflict, and the integrity level of $l_2$ is higher than that of $l_1$.
	\item The security label corresponding to an NMI at the root level, $\{[\perp, \perp, \ldots, \perp], [\omega_q]\}$, is \textbf{dominated by} all other levels.
	\item The system high, denoted by $\{[T, T, \ldots, T], [\omega_1]\}$, \textbf{dominates} all other levels.
	\item The dominance relation defines a lattice structure, where the NMI label appears at the bottom and the level trusted appears at the top. Incomparable levels are not connected in this lattice structure.
\end{itemize}

In accordance with our proposed access control model, the rules for
information flow as they apply to it are as follows:

\begin{enumerate}
	\item Simply Property: A calibration technician (S), may read a calibration report (O), only if the label, $L(S) \geq L(O)$.
	\item * (Star) Confinement Property: A calibration technician (S) can only calibrate (write) a system component or unit (O), if the label of the component dominates that of the technician, i.e. if $L(O) \geq L(S)$. Specifically, write corresponds to the production of a calibration report. 
\end{enumerate}

%% file: sections/evaluation.tex
\section{Evaluation}
\label{sec:evaluation}

While our unified model may be theoretically sound, it is vital to
evaluate whether or not such a model is practically efficient for
enforcing constraints in a real application. In this section, we first
describe a case example for which our model could be applied, and follow
on to evaluate the performance of our model as a practical solution for
authorising traceability verification checks.

\subsection{Case Example: Calibration Traceability for a Sensor Device}

To describe a case as to how our proposed model can be applied,
we will take the example of an infrared thermometer sensor. To keep
in scope of this paper, we will discount the uncertainty
calculations that make part of the traceability verification process,
as the primary concern is with the authorisation time for enforcing
our model rather than the overall time to complete calibration itself.

\begin{figure}
  \centering
  \includegraphics[width=0.75\linewidth]{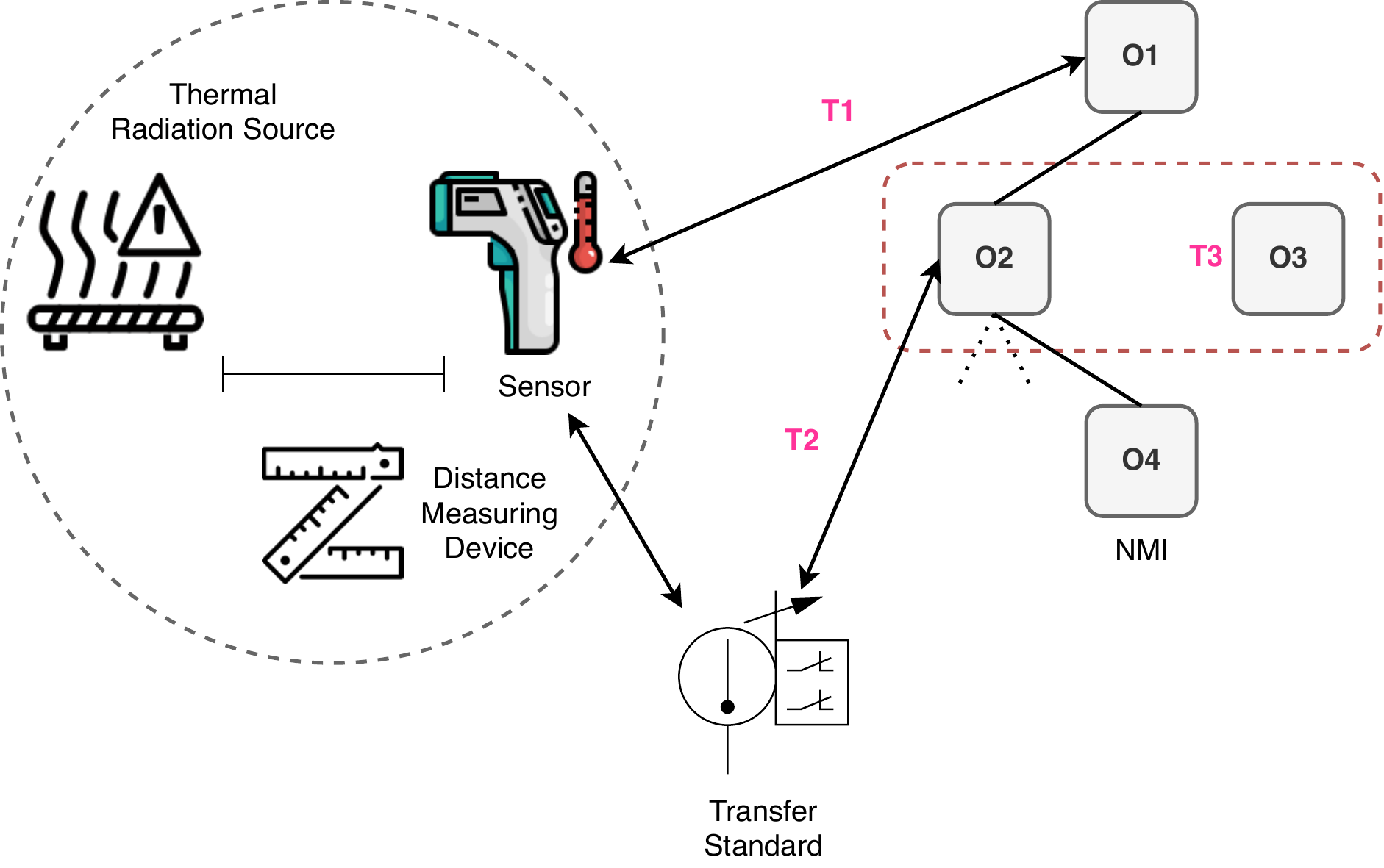}
  \caption{Traceability Chain for Infrared Thermometer}
  \label{fig:thermtrace}
\end{figure}

In order to calibrate an infrared thermometer, we need: (1) a thermal
radiation source, (2) a transfer standard (an intermediate device
when comparing other devices in calibration), (3) an ambient temperature
thermometer, and (4) a distance measuring device. In some cases, for
example where an aperture is part of its calibration, additional
equipment may be required, however we will consider the more general case
described. Furthermore, uncertainty calculations make up part of the
traceability verification process, but as we are more concerned with
the authorisation time for our model, we will not take these into account.
As shown in Figure~\ref{fig:thermtrace}, we can see an example scheme for
tracing the infrared thermometer to national standards.

To discuss our model in this case setting, we will assume the following:
\begin{itemize}
	\item The calibration facilities $O1-3$ are classed as intermediary
	      calibration facilities, and $O4$ is a National Measurement
		  Institute (NMI)
	\item The infrared thermometer sensor is calibrated by technician
	      $T1$ at calibration facility $O1$
	\item The transfer standard used in calibrating the sensor is itself,
	      calibrated by a technician $T2$ at organisation $O2$
	\item The facilities $O2$ and $O3$ are in direct competition with
	      one another
	\item The traceability chain information flows from the sensing
	      device to $O1$, to the transfer standard calibrated by a
		  technician at $O2$, towards the NMI $O4$
\end{itemize}

Given the assumptions above, we now describe the calibration traceability
lifecycle for our infrared thermometer sensor, and how our proposed
unified model can be applied.

\paragraphb{Initial Calibration}

Inspired by biology, we refer to the initial calibration of a system
component or test and measurement equipment as the \textit{birth} of
the device, representing the behaviour which implements its secure
initialisation. For all devices, there is always an initial calibration
step which commences the start of its lifecycle and inaugurates them
into the calibration ecosystem. After manufacturing, there are three
primary methods of initial calibration: (a) by the manufacturer; (b)
by a third-party intermediary calibration facility; and (c) at a
National Measurement Institute (NMI).

In accordance with our model, and taking our case example, the sensor's
initial calibration will be conducted by the technician $T1$ at the
facility $O1$. This process will output a calibration report for the
sensor and this will be given the security label assigned to the
technician: $\{[\perp, \perp, \perp], [\omega_1]\}$.
Similarly, the transfer standard is calibrated by the technician $T2$
at facility $O2$ and will have the security label:
$\{[\perp, COI_1, \perp], [\omega_2]\}$, where $COI_1 = \{O2, O3\}$
denotes that facilities $O2$ and $O3$ are in conflict with each other.

In most cases, there will be no conflicts that arise as part of a
device's initial calibration, however some devices such as those created
for military or other government organisations may be classified
in nature. Thus, if the organisation in this case previously used a
calibration facility for calibrating a set of other devices, the
new facility to be contracted could be competition/conflict with the
other and thus the labels would indicate this conflict in the chain.

\paragraphb{Verifying Traceability}

Traceability plays a key role in verifying a newly calibrated device,
as well as determining whether it needs to be recalibrated; ultimately
lying at the heart of on-the-fly calibration. As the traceability
verification process involves first retrieving the report of the device
we are tracing, and then its parents' reports and so forth to the root
level, the party who carries out the verification check must first
satisfy a set of conditions before being allowed to do so.

In accordance with our access control model, the security label of the
device being traced must dominate the label of the party carrying out
the verification check. For example, in the case of recalibration which
first involves a traceability check, the label for a device must
dominate that of the verifier. Furthermore, as we require to read
all the reports of all parents at each step in the chain,
up to national standards, the label of each parent should also
dominate that of the verifying party, such that for a traceability
chain $C = \{L(O_1), L(O_2), \ldots, L(O_n)\}$, $\forall c \in C$, $L(S) \geq c_i$,
where $i,n \geq 1$ and $L(S)$ is the label of the party carrying out
the verification.

\paragraphb{Recalibration}

The process of recalibration is one that is carried out either: (a) when the expiry date of a component/unit's calibration report has been exceeded, (b) when critical measurements are taken, (c) if the accuracy or measurement uncertainty of the equipment has noticeably degraded or drifted before the expiry period, or (d) if any parent unit in the chain does not have valid calibration. In any case, recalibration first involves performing a traceability verification check, to ensure that the component/unit has unbroken traceability and valid calibration. The technician performing calibration must first be allowed to carry out the traceability check, such that the label of the component or unit being calibrated dominates that of the technician, $L(O) \geq L(S)$. Specifically, this dominance relation is satisfied when the technician is not in conflict; i.e. if the technician is the same as the one who performed the initial calibration or previous recalibration, then they will be allowed to do so. Similarly, if not then the model would verify that the new technician performing recalibration is not in conflict with the previous, or others in the chain. That is, the traceability chain of the equipment being recalibrated does not contain information from both technicians in a conflict of interest set $COI_j$.

%
%


\subsection{Performance Evaluation}

As well as discussing how the model will be applied in a real-world
case, it is vital to determine its practicality. We mention that the
individual access control models, or a simple conjunction of them, may
not be suitable or efficient compared to the unified model. In this
evaluation, we determine whether our unified model can be enforced in
a practical setting to support efficient calibration verification
\textit{on-the-fly} and scale well with large, complex calibration
hierarchies.

\paragraphb{Setup}

\begin{figure}
  \centering
  \includegraphics[width=0.5\linewidth]{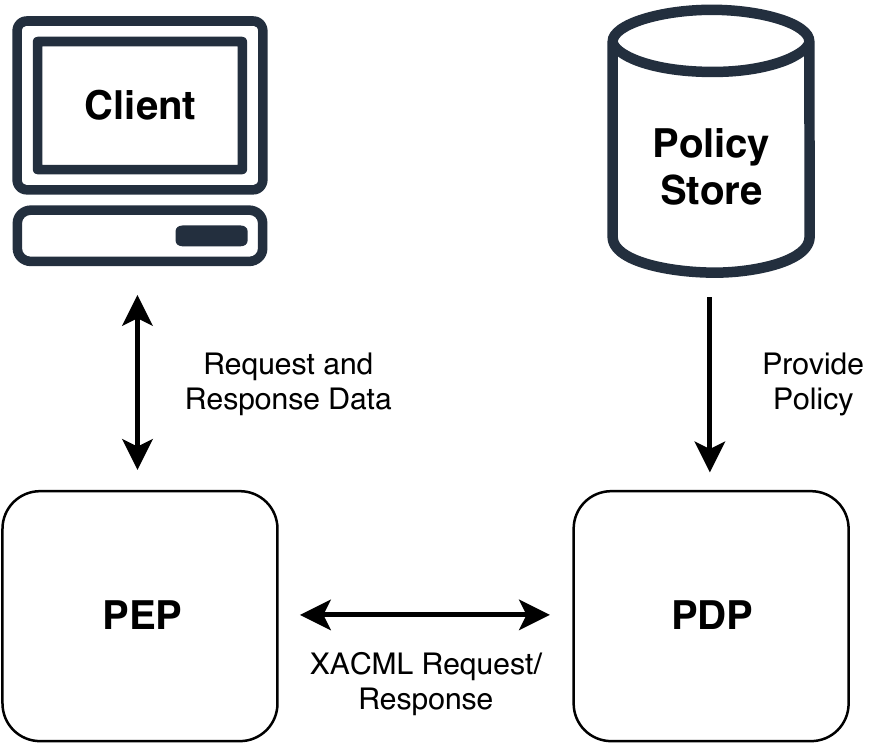}
  \caption{XACML Authorisation Architecture}
  \label{fig:xacmlarch}
\end{figure}

To evaluate our model, we made use of an attribute-based authorisation
framework, following the XACML standard~\cite{oasisxacmldoc} and
structured as shown in Figure~\ref{fig:xacmlarch}. The framework provides
standards for access requests and policy specification, where a client
program acts as a Policy Enforcement Point (PEP) which sent access requests
to and enforced responses given by the server running a Policy Decision
Point (PDP). The experiments on the authorisation framework to evaluate
our model were performed on a virtual machine running Ubuntu 14.04 LTS
allocated with 64GB of RAM.

\paragraphb{Baseline Model}
To provide a more in-depth comparison as to how our access control model
performs, we conducted the evaluation starting with a \textit{simple
conjunction} of the three models (BLP, Reverse BIBA (RBIBA) and
Chinese Walls), as a \textbf{baseline model} to compare our unified model
against. Specifically, to create the baseline model we made use
of the XACML {\em PolicySet}, where policies for each model can be
combined into a single policy and enforced together using a
{\em PolicyCombiningAlgorithm}. To combine the policies for our
baseline, we used the {\em permit-unless-deny} algorithm, which only
allows a Permit or Deny response, and will deny access if any one of
the combined policies produce a Deny result.

\paragraphb{Authorising Traceability Verification}

For the first part of our evaluation, we measured the time taken to
authorise access requests for calibration traceability verification
using our unified model. As a point of comparison, we also measure
this time for the baseline model. To recall, traceability verification
involves verifying the calibration at each level in the chain of
calibrations for some equipment, up to national standards.

Naturally, in the calibration hierarchy, there are several levels for
a single traceability chain. For example, a simple temperature sensor
could be calibrated with a platinum resistance thermometer, which in
turn is calibrated by a more accurate reference thermometer, that is
finally calibrated by a helium gas thermometer (primary reference
standard)~\cite{morris2012measurement}. However, with the consideration
of SC-IoT systems, where sensors, etc. could be off-the-shelf
components, possibly leading to longer traceability chains (>4 levels).
Thus, for completeness, we conduct our experiment for measuring
the authorisation time for traceability verification with up to 50
levels.

\begin{figure}[h]
  \centering
  \includegraphics[width=0.5\linewidth]{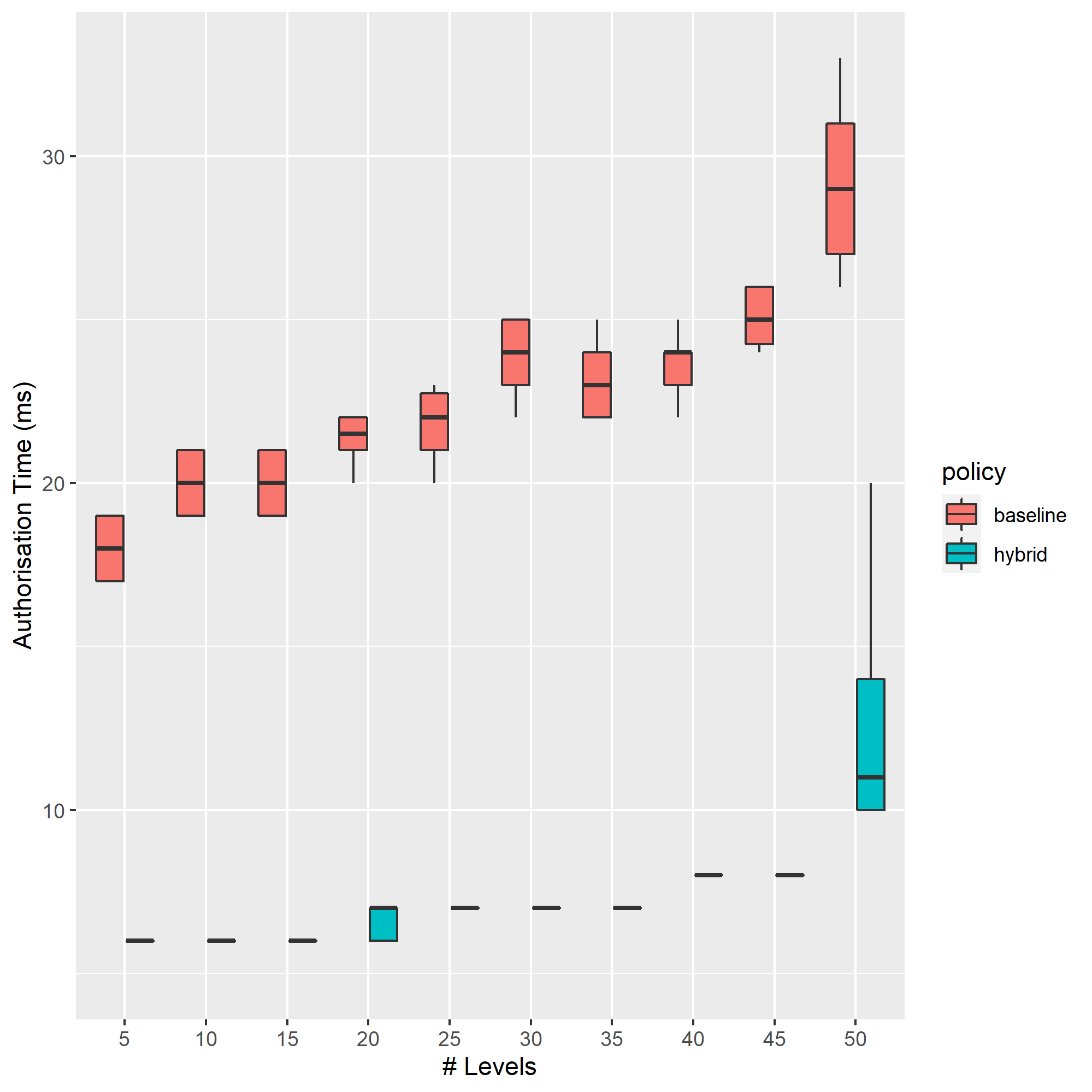}
  \caption{Authorisation Time for Single Traceability Chain}
  \label{fig:verify_1_branch_bplots}
\end{figure}

For the first experiment, we measured the time to authorise
traceability verification for up to 50 levels, with only a single
parent (reference device) at each level. As shown in
Figure~\ref{fig:verify_1_branch_bplots}, we observe that across the
board, the unified model is significantly faster compared to the
baseline simple conjunction, with authorisation times not exceeding
11ms on average in the worst case compared to roughly 30ms for the
baseline model.

\begin{figure}[h]
\centering
\subfigure[2 Branches Per Level]{
\includegraphics[width=.45\textwidth]{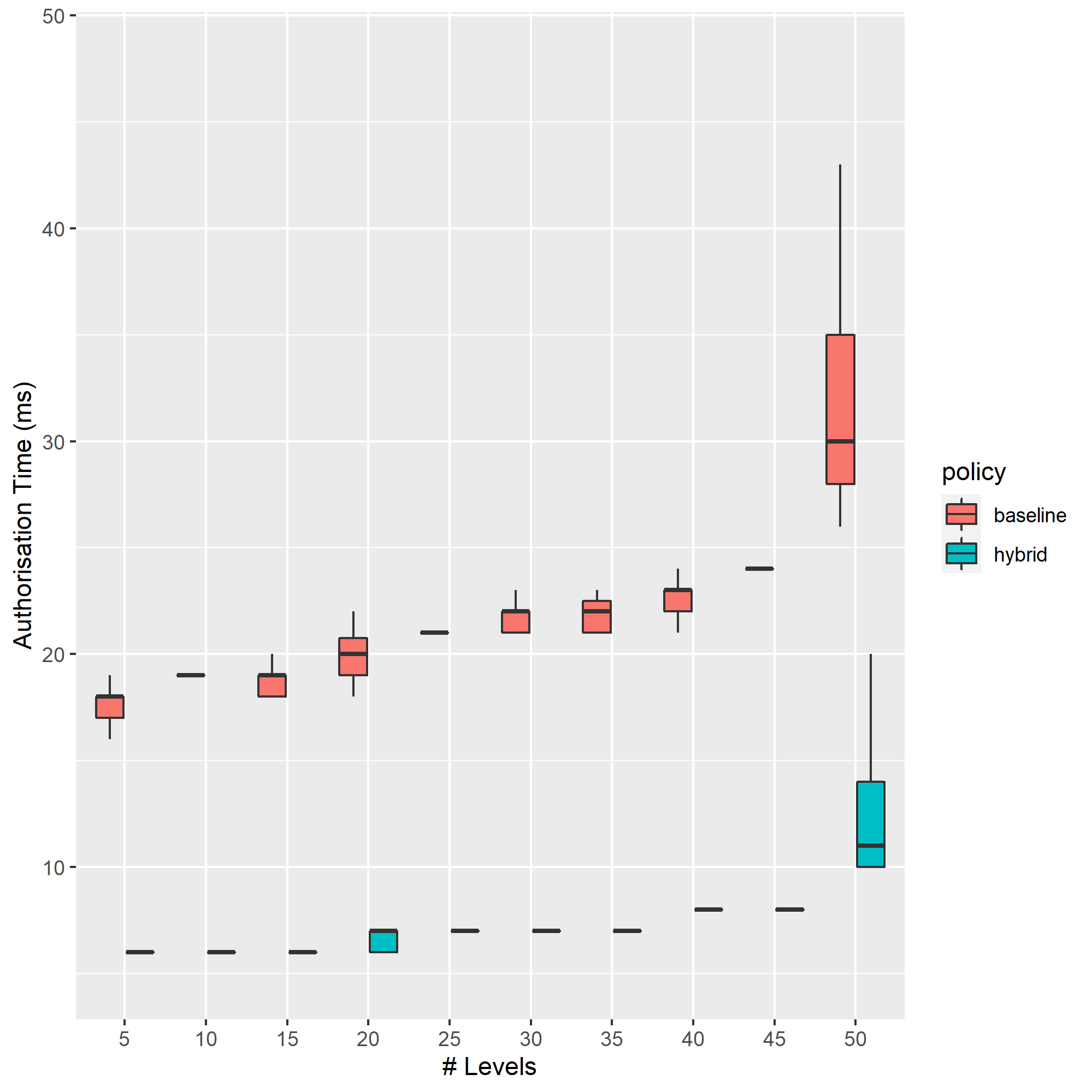}
\label{fig:verify_2_branch_bplots}
}
\subfigure[4 Branches Per Level]{
\includegraphics[width=.45\textwidth]{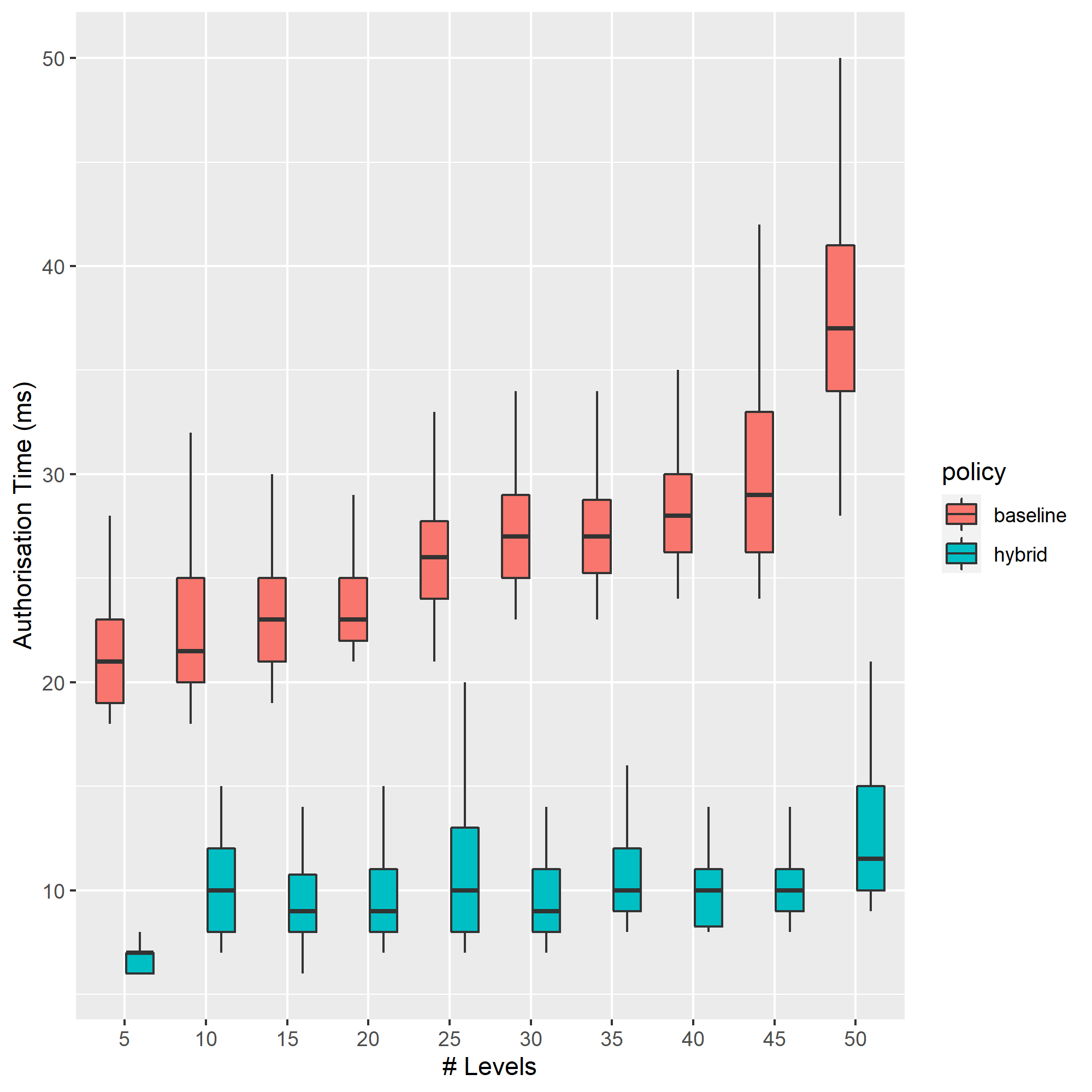}
\label{fig:verify_4_branch_bplots}
}
\caption{Authorisation Times for Branching Traceability Chains}
\end{figure}

In realistic traceability chains, some devices are calibrated with
more than one reference (parent) device. For example, an infrared
temperature sensor is calibrated with a distance gauge, infrared source
(i.e. hotplates) and reference thermometer~\cite{liebmann2011infrared}. Each of these parent
devices will have their own parents, meaning that instead of having
a single chain, we now have a chain that branches off several times.
Therefore, our next experiment involves measuring the authorisation
time for traceability verification for 2 and 4 branches per level,
such that we can observe the impact of realistically complex chains.
As shown in Figures~\ref{fig:verify_2_branch_bplots} and
\ref{fig:verify_4_branch_bplots}, we see a similar pattern to a single
branch for a chain, with our unified model outperforming the baseline
model by at least 15ms in the worst case. Furthermore, we can also
observe an increase in the authorisation time as the number of branches
per level also increases, as authorisations for each parent device
and so forth in their branches also need to be made.

Across all test cases, while the unified model does
outperform (significantly) the baseline simple conjunction, the authorisation
time does increase as the number of levels increases. However, in the
case of our unified model, the increase of just a few milliseconds as
the number of levels or branches increase is considered a reasonable
wait on the device prep-cycle or if a measurement was to
be taken~\cite{chu2019respiration}.


%

\paragraphb{Conflict Management}

Pertaining to the conflict component of security labels, we note that it is in fact the size of conflict sets, rather than the number of them, which needs to be considered, as some sets may only contain two members, whilst others could contain more. Thus, we evaluated the effect of the size of conflict sets on authorisation time in traceability verification, for our unified model.

For our calibration traceability dataset, we generated a set of calibration reports
which contained real calibration data. Each of which were assigned a
security label, where the integrity component of the label corresponded
with the level at which calibration was conducted, and the conflict
component of the label was generated using a $G(n, p)$ variant of the
Erdős–Rényi random graph model, where $n$ is the number of potential
competitors and $p$ is the probability of conflict. Specifically, the
cliques of the random graph represented conflict sets, which for each
node $n_i$ was assigned a set of conflict sets.

\begin{figure}[h]
  \centering
  \includegraphics[width=0.5\linewidth]{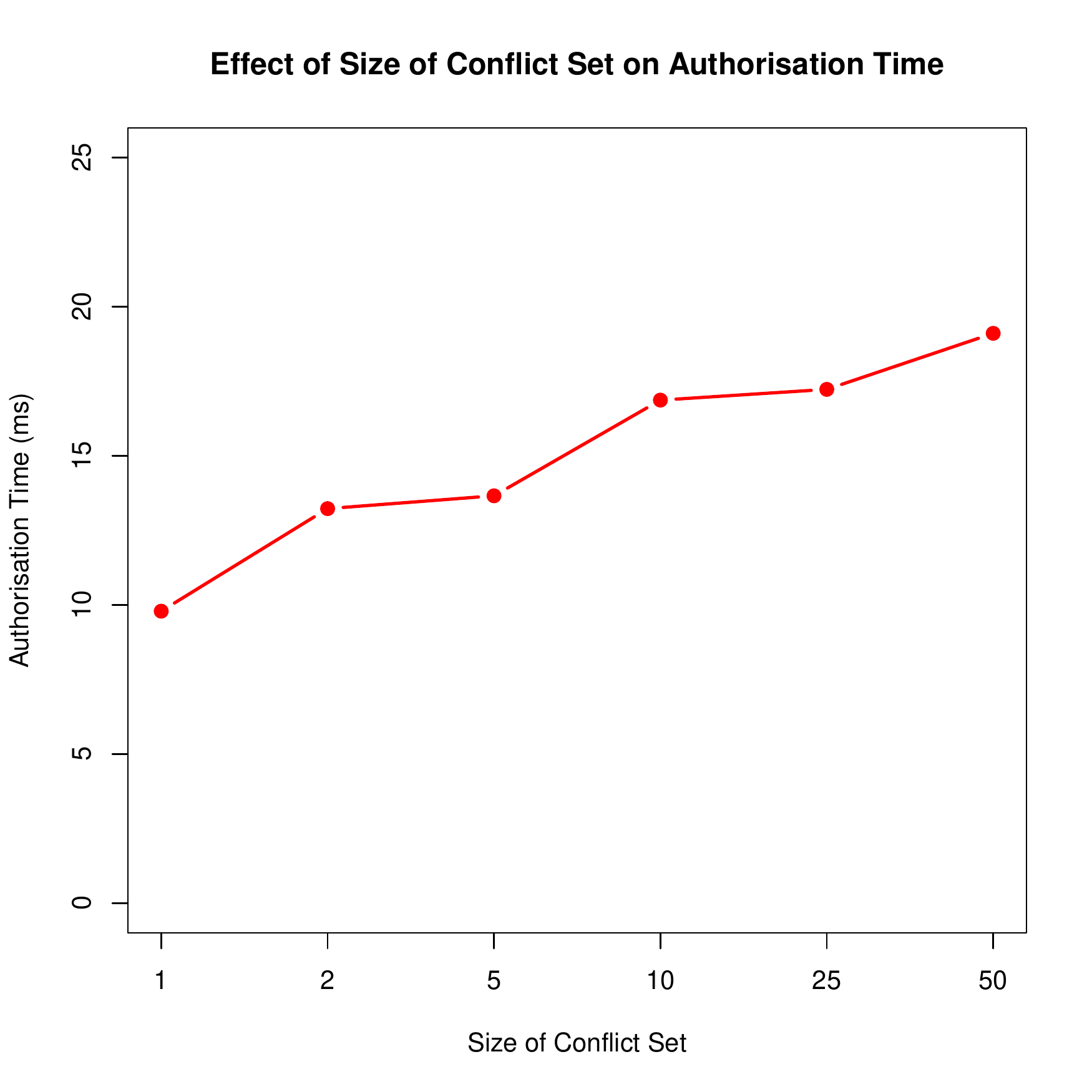}
  \caption{Effect of Conflict Set Size on Authorisation Time}
  \label{fig:conflict_size_time}
\end{figure}

To evaluate the impact of the size of conflict sets, we increase the
number of potential competitors $n$ and probability of potential
competitors being in conflict $p$, such that we get a range of conflict
set sizes from 1 member in the set to 50. As shown in
Figure~\ref{fig:conflict_size_time}, we observe an increase in the
time taken to authorise traceability verification requests as the
size of the conflict sets increase. Our experiment here shows the
effect on authorisation time for a single conflict set, however, we
can see clearly that as the number of conflict sets increase, so will
the authorisation time; but impressive yet the increase is fairly
minimal and as previously stated is considered a reasonable
wait on the device prep-cycle.

%% file: sections/discussion.tex
\section{Discussion and Limitations}
\label{sec:discussion}


Within the calibration hierarchy, we observe a unique set of information
flows which we hypothesised could not be effectively met by classical
access control models, as any one of them or a simple conjunction of a
subset such as the lattice model, fails to meet the desired access
control requirements for compartmentalising conflicts of interest,
calibration integrity and confidentiality.

With some key players (device operators, calibration facilities, etc.)
in the calibration ecosystem sharing an adversarial relationship with
a subset of others, we observe a unique set of information flows. First,
we observe that calibration traceability checks could reveal
\textit{who-calibrates-for-whom}, which can ultimately compromise the
confidentiality of business relationships surrounding SC-IoT systems.
With the adoption of our unified model, we constrain these information
flows such that the field level where systems are deployed
retain the highest level of confidentiality and calibration
meta-information such as \textit{what-is-being-calibrated} and
\textit{how-often-it-is-calibrated} cannot be revealed to levels above
the field level (i.e. intermediary facilities).

One limitation of this work, is that policies and calibration reports are stored centrally. We assume that encryption mechanisms are in place, however we do not consider the impact of security mechanisms such as TLS and PKI on authorisation times in our evaluation. In any case, added security presents overheads that would need to be considered, and in certain cases (i.e. critical liabilities resulting from calibration checks before system operation can continue) the addition of such security measures could increase the low authorisation times
observed from evaluation of our unified model.

%% file: sections/related_work.tex
\section{Related Work}
\label{sec:related}

The security surrounding time- and safety-critical IoT systems is an
active research area~\cite{chen2018securing}, with the main focus
pertaining to attacks in the cyber domain (i.e. control system
security~\cite{hasan2019protecting,alemzadeh2016targeted}) and the
physical domain (i.e. physical compromise of sensors~\cite{ali2018cyber,chowdhary2017security} and
physical safety of devices and surrounding environment~\cite{salmi2018human,chowdhary2017security}). However,
with the calibration of such systems contributing highly to system
accuracy and precision, the compromise of its calibration can
ultimately impact the ability to operate safely. Quarta et al.
describe calibration parameters of robotics systems to be an essential
construct used to compensate for known measurement
errors~\cite{quarta2017experimental}. They demonstrated that by
manipulating calibration parameters, an adversary could cause the robot
to operate unsafely, such as affecting the servo motor causing the
robot to move erratically. Aside from just the manipulation of calibration
parameters, we observe a set of calibration information flows,
possibly between calibration parties that share conflicts-of-interest,
which can lead to compromise of confidentiality and integrity of
calibration.


Existing access control literature for safety-critical IoT systems focus
on various aspects of the system itself, ranging from securing control
systems (i.e. access to actuators) to the validation/enforcement of
security policies. Hasan and Mohan~\cite{hasan2019protecting} propose a framework
based on the Simplex architecture, commonly used for time-critical cyberphysical
systems for fault-tolerance, which makes use of a rule-based invariant
and access control mechanism to ensure the timing and safety requirements
of IoT cyberphysical systems (i.e. ensure some task can only access a
given actuator if the task has the required permission given a set of
invariant conditions). Frank et al.~\cite{frank2004combining} describe a
combination of both logical and physical access control -- explain each,
respectively. He describes that the most widely used multi-level security
models are inadequate when logical resources obtain a physical form, that makes
use of both mandatory and discretionary access control. While not directly applicable
to calibration traceability, one must also consider the access constraints to
the physical process of calibration which traceability verification is a key part
of. Compared to our work, we make use of attribute-based access control (ABAC)
due to its flexibility, limited only by the computational language when
implementing policies to enforce access control models. Specifically, it allows
greater breadth for access relationships (subjects to access objects) without
the need to specify the individual relationships between them. Compared to
other traditional approaches such as rule-based access control, this makes its
use ideal for dynamic environments such as SC-IoT~\cite{hu2015attribute}.
Further, while attribute-based access control provides key advantages to earlier
forms of access control, such as ACLs and role-based access control,
and having a well-maintained policy declaration language and
authorisation framework (XACML) for practical solutions, other forms
of access control have been proposed which may also be suitable for
enforcing our unified model.

In this work, we focus on the calibration angle which has been paid little
attention to. Specifically, we focus on the unification of three classical
access control models (namely BLP, BIBA and Chinese Walls) which is required to
solve the novel set of information flows which arise from calibration traceability.
Yang et al.~\cite{yang2016improved} state that while BLP is widely used to enforce
multi-level confidentiality, it lacks flexibility due to strict
confidentiality rules. Furthermore, they describe that BLP poorly
controls integrity and that BLP is commonly combined with BIBA for
increased integrity control~\cite{shi2001research,he2007analysis,zhang2008research}. In their work, they propose an improved
BLP model to manage multi-level security, where the security
of each level is distinguished by the security level of the accessed
content itself (subjects are defined as a multi-level entity and objects
are defined as a single-level entity).
With regard to BIBA, Liu et al.~\cite{liu2017btg} note that BIBA can
possibly deny non-malicious access requests made by subjects, ultimately
reducing the availability to a system. To this, they propose the
integration of notions from Break The Glass (BTG) strategies --
a set of (efficient) strategies used to extend subject access rights in
exceptional cases (i.e. irregular system states) -- with the existing BIBA
model (BTG-BIBA). They show that with the proposed BTG-BIBA model,
it can now provide more fine-grained access control that is context-aware
for dynamic situations. In this work, we take into account the traditional
BIBA and BLP models in our case for unification, however one can question
the applicability of improvements made to these models over recent years.
For example, capability-based access control has been shown to perform
well in highly scalable and distributed environments, such as IoT.
Similarly, like attribute-based access control, capability-based
mechanisms can also be enforced in a fine-grained
manner~\cite{anggorojati2012capability,gusmeroli2013capability}, where
tokens can be given to subjects on-the-fly containing the appropriate
security label, and also be verifiable and unlinkable to preserve
privacy~\cite{zhang2009dp2ac}.

%% file: sections/conclusion.tex
\section{Conclusion}
\label{sec:conclusion}

We have highlighted the shift towards a digital calibration paradigm presents
us with a novel access control challenge when we consider the calibration of
rapidly adopted safety-critical IoT systems.Upon discussion of the current
state-of-the-art in calibration traceability, we observe the information flow
through a systems calibration hierarchy and present an access control model which
uniquely unifies the BLP, BIBA and Chinese Wall models. Furthermore, we have
developed an authorisation framework to evaluate the performance of our model
for safety-critical IoT systems, and have shown that authorisation times can
suitably enforce restrictions that enable efficient, safe calibration traceability.


%% file: paper.bbl
\begin{thebibliography}{10}
\providecommand{\url}[1]{\texttt{#1}}
\providecommand{\urlprefix}{URL }
\providecommand{\doi}[1]{https://doi.org/#1}

\bibitem{abomhara2014security}
Abomhara, M., K{\o}ien, G.M.: Security and privacy in the internet of things:
  Current status and open issues. In: 2014 international conference on privacy
  and security in mobile systems (PRISMS). pp.~1--8. IEEE (2014)

\bibitem{alemzadeh2016targeted}
Alemzadeh, H., Chen, D., Li, X., Kesavadas, T., Kalbarczyk, Z.T., Iyer, R.K.:
  Targeted attacks on teleoperated surgical robots: Dynamic model-based
  detection and mitigation. In: 2016 46th Annual IEEE/IFIP International
  Conference on Dependable Systems and Networks (DSN). pp. 395--406. IEEE
  (2016)

\bibitem{ali2018cyber}
Ali, B., Awad, A.I.: Cyber and physical security vulnerability assessment for
  iot-based smart homes. sensors  \textbf{18}(3), ~817 (2018)

\bibitem{anggorojati2012capability}
Anggorojati, B., Mahalle, P.N., Prasad, N.R., Prasad, R.: Capability-based
  access control delegation model on the federated iot network. In: Wireless
  Personal Multimedia Communications (WPMC), 2012 15th International Symposium
  on. pp. 604--608. IEEE (2012)

\bibitem{bargar1998primary}
Bargar, W.L., Bauer, A., B{\"o}rner, M.: Primary and revision total hip
  replacement using the robodoc (r) system. Clinical Orthopaedics and Related
  Research (1976-2007)  \textbf{354},  82--91 (1998)

\bibitem{de2000calibration}
de~Castro, C.N., Louren{\c{c}}o, M., Sampaio, M.: Calibration of a dsc: its
  importance for the traceability and uncertainty of thermal measurements.
  Thermochimica Acta  \textbf{347}(1-2),  85--91 (2000)

\bibitem{chen2018securing}
Chen, C.Y., Hasan, M., Mohan, S.: Securing real-time internet-of-things.
  Sensors  \textbf{18}(12), ~4356 (2018)

\bibitem{chowdhary2017security}
Chowdhary, S., Som, S., Tuli, V., Khatri, S.K.: Security solutions for physical
  layer of iot. In: 2017 International Conference on Infocom Technologies and
  Unmanned Systems (Trends and Future Directions)(ICTUS). pp. 579--583. IEEE
  (2017)

\bibitem{chu2019respiration}
Chu, M., Nguyen, T., Pandey, V., Zhou, Y., Pham, H.N., Bar-Yoseph, R.,
  Radom-Aizik, S., Jain, R., Cooper, D.M., Khine, M.: Respiration rate and
  volume measurements using wearable strain sensors. NPJ digital medicine
  \textbf{2}(1), ~1--9 (2019)

\bibitem{dabbagh2019internet}
Dabbagh, M., Rayes, A.: Internet of things security and privacy. In: Internet
  of Things from hype to reality, pp. 211--238. Springer (2019)

\bibitem{frank2004combining}
Frank, K., Willemoes-Wissing, I.C.: Combining logical and physical access
  control for smart environments. Master's thesis, Technical University of
  Denmark, DTU, DK-2800 Kgs. Lyngby, Denmark (2004)

\bibitem{gusmeroli2013capability}
Gusmeroli, S., Piccione, S., Rotondi, D.: A capability-based security approach
  to manage access control in the internet of things. Mathematical and Computer
  Modelling  \textbf{58}(5-6),  1189--1205 (2013)

\bibitem{hasan2019protecting}
Hasan, M., Mohan, S.: Protecting actuators in safety-critical iot systems from
  control spoofing attacks. In: Proceedings of the 2nd International ACM
  Workshop on Security and Privacy for the Internet-of-Things. pp. 8--14 (2019)

\bibitem{he2007analysis}
He, J.B., Qing, S.H., Wang, C.: Analysis of two improved blp models. Ruan Jian
  Xue Bao(Journal of Software)  \textbf{18}(6),  1501--1509 (2007)

\bibitem{hu2015attribute}
Hu, V.C., Kuhn, D.R., Ferraiolo, D.F., Voas, J.: Attribute-based access
  control. Computer  \textbf{48}(2),  85--88 (2015)

\bibitem{hwang2015iot}
Hwang, Y.H.: Iot security \& privacy: threats and challenges. In: Proceedings
  of the 1st ACM workshop on IoT privacy, trust, and security. pp.~1--1 (2015)

\bibitem{liebmann2011infrared}
Liebmann, F.: Infrared thermometer calibration. Cal Lab Int. J. Metrol pp.
  20--22 (2011)

\bibitem{liu2017btg}
Liu, G., Wang, C., Zhang, R., Wang, Q., Song, H., Ji, S.: Btg-biba: A
  flexibility-enhanced biba model using btg strategies for operating system.
  International Journal of Computer and Information Engineering
  \textbf{11}(6),  765--771 (2017)

\bibitem{morris2012measurement}
Morris, A.S., Langari, R.: Measurement and instrumentation: theory and
  application. Academic Press (2012)

\bibitem{oasisxacmldoc}
OASIS: extensible access control markup language (xacml) version 3.0,
  \url{\url{https://docs.oasis-open.org/xacml/3.0/xacml-3.0-core-spec-os-en.html}}

\bibitem{quarta2017experimental}
Quarta, D., Pogliani, M., Polino, M., Maggi, F., Zanchettin, A.M., Zanero, S.:
  An experimental security analysis of an industrial robot controller. In: 2017
  IEEE Symposium on Security and Privacy (SP). pp. 268--286. IEEE (2017)

\bibitem{salmi2018human}
Salmi, T., Ahola, J.M., Heikkil{\"a}, T., Kilpel{\"a}inen, P., Malm, T.:
  Human-robot collaboration and sensor-based robots in industrial applications
  and construction. In: Robotic Building, pp. 25--52. Springer (2018)

\bibitem{sandhu1993lattice}
Sandhu, R.S.: Lattice-based access control models. Computer (11),  9--19 (1993)

\bibitem{shi2001research}
Shi, W.: Research on and enforcement of methods ofmethods of secure operating
  systems development [ph. d. thesis]. Institute of Software, The Chinese
  Academy of Sciences, Beijing  (2001)

\bibitem{vim2004international}
Vim, I.: International vocabulary of basic and general terms in metrology
  (vim). International Organization  \textbf{2004},  09--14 (2004)

\bibitem{wang2010security}
Wang, E.K., Ye, Y., Xu, X., Yiu, S.M., Hui, L.C.K., Chow, K.P.: Security issues
  and challenges for cyber physical system. In: 2010 IEEE/ACM Int'l Conference
  on Green Computing and Communications \& Int'l Conference on Cyber, Physical
  and Social Computing. pp. 733--738. IEEE (2010)

\bibitem{yaugdereli2015study}
Ya{\u{g}}dereli, E., Gemci, C., Akta{\c{s}}, A.Z.: A study on cyber-security of
  autonomous and unmanned vehicles. The Journal of Defense Modeling and
  Simulation  \textbf{12}(4),  369--381 (2015)

\bibitem{yang2016improved}
Yang, P., Wang, Q., Mi, X., Li, J.: An improved blp model with more
  flexibility. In: 2016 13th International Conference on Embedded Software and
  Systems (ICESS). pp. 192--197. IEEE (2016)

\bibitem{zhang2008research}
Zhang, J., Yun, L.J., Zhou, Z.: Research of blp and biba dynamic union model
  based on check domain. In: 2008 International Conference on Machine Learning
  and Cybernetics. vol.~7, pp. 3679--3683. IEEE (2008)

\bibitem{zhang2009dp2ac}
Zhang, R., Zhang, Y., Ren, K.: Dp$^2$ac: Distributed privacy-preserving access
  control in sensor networks. In: IEEE INFOCOM 2009. pp. 1251--1259. IEEE
  (2009)

\end{thebibliography}
